\documentclass[10pt]{article}
	\usepackage{latexsym,graphicx,multirow}
	\usepackage{amssymb}
	\usepackage{amsmath}
	\usepackage{amscd}
	\usepackage{amsthm}
	\usepackage[left=2cm,top=2.5cm,right=2.5cm,bottom=1.5cm]{geometry}
	\usepackage{hyperref}
	\usepackage{epstopdf}
	\usepackage{float}
	
\begin{document}
		
	\begin{center}
	\large{\bf{Stability, dark energy parameterization and swampland aspect of Bianchi Type-$ VI_{h}$ cosmological models 
	with f(R, T)-gravity}} \\
	\vspace{10mm}
	\normalsize{ Archana Dixit$^1$ and Anirudh Pradhan$^2$  }\\
	\vspace{5mm}
	\normalsize{$^{1,2}$Department of Mathematics, Institute of Applied Sciences and Humanities, GLA University\\
			Mathura-281 406, Uttar Pradesh, India}\\
	\vspace{2mm}
	$^1$E-mail: archana.dixit@gla.ac.in.\\
	$^2$E-mail: pradhan.anirudh@gmail.com\\
	
		\vspace{10mm}
	\end{center}
	\begin{abstract}
	
	Stability, dark energy (DE) parameterization and swampland aspects for the Bianchi form-$VI_{h}$ universe have been formulated 
	in an extended gravity hypothesis. Here we have assumed a minimally coupled geometry field with a rescaled function of $f (R, T)$ 
	replaced in the geometric action by the Ricci scalar $ R$. Exact solutions are sought under certain basic conditions for the related 
	field equations. For the following theoretically valid premises, the field equations in this scalar-tensor theory have been solved. 
	It is observed under appropriate conditions that our model shows a decelerating to accelerating phase transition property. Results are 
	observed to be coherent with recent observations. Here, our models predict that the universe's rate of expansion will increase with the 
	passage of time. The physical and geometric aspects of the models are discussed in detail. In this model, we also analyze the 
	parameterizations of dark energy by fitting the EoS parameter $\omega(z)$ with redshift. The results obtained would be useful in 
	clarifying the relationship between dark energy parameters. In this, we also explore the correspondence of quintessence dark energy 
	with swampland criteria. The swampland criteria have  been also shown the nature of the scalar field and the potential of the scalar field.
	\end{abstract}
	 \smallskip
 
 
\section{Introduction}
Our view of cosmology has been revolutionized by recent cosmological observations.
This appears that there is an exponential expansion of the currently observable
universe \cite{ref1}-\cite{ref3}. The source that drives this acceleration is termed as `dark energy', 
whose beginning in present-day cosmology is still a mystery. This is a result of the way that we do not have, 
up until now, a predictable hypothesis of quantum gravity. New prove from astronomy and cosmology has
recently revealed a rather surprising image of the Universe. Our new datasets from different sources, for example, 
Cosmic Microwave Background Radiation(CMBR) and Supernovae, imply that the universe's vitality spending plan resembles as: 
$4\%$ typical baryonic matter, $22\%$ dark matter, and $74\%$ dark energy \cite{ref4}-\cite{ref7}.\\
One of the hot topic issues between cosmologists and astrophysicists is the accelerated problem of expansion of the universe. 
The common view suggests that it is dark energy and updated 
alternative theories to explain the expansion of the universe.
Big alternative theories are $f(R)$ \cite{ref8} theory, $f(G)$ \cite{ref9} theory,
$f(R, T)$ \cite{ref10} theory, and so on. Author \cite{ref10} proposes the concept 
of $f (R, T)$ to describe accelerated universe expansion. 
They come up with three models to overcome the argument.
They also constructed a remodelled theory of F(R, T) gravity
in which the gravitational Lagrangian is described by an arbitrary function of
The Ricci scalar $R$ and the stress energy tensor trace $T$. Bianchi type
models with anisotropic spatial segments are fascinating as in they are
more broad than the Friedman models. Despite the fact that there is a solid discussion 
going on the reasonability of Bianchi type models \cite{ref11}, these models can be helpful in the portrayal of early inflationary 
stage and with appropriate component can be diminished to isotropic conduct at late occasions. In the system of f(R) gravity, numerous 
authors have examined various parts of Bianchi type models \cite{ref12}-\cite{ref14}. As of late anisotropic cosmological 
arrangements in $f(R)$ gravity have gotten in \cite{ref15}.\\

Bianchi type models are  anisotropic and homogeneous. Such models are nine altogether, yet their grouping 
permits them to be divided into two classes. Bianchi ($I$, $II$, $VII$, $VIII$ and $IX$) models are in class $A$ and 
Bianchi ($III$, $IV$, $V$, $VI$ and $VII$) are in class $B$. Spatially homogeneous cosmological models assume a significant 
job in clarifying the structure and space properties of all Einstein field conditions and cosmological arrangements. 
Bianchi type-$I$ axially symmetrical cosmological models with a magnetic field investigated \cite{ref16} and string 
cosmology in Bianchi $III$ and $ VI_{0}$ has discussed in \cite{ref17}-\cite{ref18}. Bianchi type-III cosmological 
model in $f (R, T)$ hypothesis of gravity have acquired in \cite{ref19}, Bianchi types-$I$ and $V$ cosmological models in 
$f(R, T)$ gravity have acquired careful arrangements of \cite{ref20}  and  \cite{ref21} have considered another class 
of cosmological models in $f(R, T)$ gravity. Recently, Bianchi type $VI_{h}$ universe model in $f(R,T)$ gravity has been 
studied by\cite{ref22}. There are not many studies in modified theories in the literature on the Bianchi form $VI_{h}$ 
metric. Then in this analysis we investigated the distribution of MSQM in $f(R, T)$ gravitation theory for Bianchi $VI_{h}$ 
universe \cite{ref23}. Many experiments have been carried out in this theory to study the dynamical aspects of anisotropic 
cosmological models \cite{ref24}-\cite{ref28}.\\

Bianchi type $ VI_{h}$ cosmological model with perfect fluid as an origin of gravitational field is given in the theory 
of self-creation for various cases of matter discussed by \cite{ref29}-\cite{ref30}. This addresses the model's physical and
mathematical properties. The objective of the paper is to study the Bianchi type $ VI_{h}$ model as suggested by Harko 
et al.\cite{ref10} as part of an extended theory of gravity. In the present work, our inspiration is to build up a general formalism 
to research these anisotropic models with a simulated time-differing deceleration parameter. In  earlier works 
\cite{ref31}-\cite{ref33}, the idea of this scale factor was already conceived. Time varying $DP$ is required to explain 
a progress of the universe from a decelerated stage to a accelerated stage at an ongoing age. The other purpose of this 
paper is to study the swampland criteria in Bianchi models. In this way, the Swampland criteria can be utilized to oblige 
the quintessence $DE$ models that begin from a scalar field theory. In literature there are lots of works that address the Swampland 
criteria and quintessence models \cite{ref34}-\cite{ref37}. Here now we investigate whether the dark energy complies with 
the swampland criteria. The string swampland criteria for a powerful field theory to be reliable with the string theory.\\

Right now explore the consistency of the dark energy with swampland measure. Also we investigate the variation of dark energy 
condition of state parameter. Here we compute the variations of $\omega$ with $z$ for dark energy model. 
A comparable test for swampland parameters for the scalar quintessence model has been performed in a prior work \cite{ref38}. 
They have been considered a core of this research scope and investigated the swampland criteria for the quintessence dark energy scope.
In doing so they have taken the trial limits by composing the varieties of dark energy condition of state in the standard CPL 
parameterization shape and afterward make an interpretation of it to acquire an upper bound of a recreated condition of $\omega (z)$. 
Here we specified two parameters  $\omega_{0 }$ and $\omega_{a}$, dark energy parameters, with the recent available SNe Ia , BAO and plank (2018) 
observation data. Chevallier-Polarski-Linder (CPL) parameterization \cite{ref39}-\cite{ref42}, Jassal-Bagla-Padmanabhan (JBP)\cite{ref43}, 
Barboza-Alcaniz parameterization \cite{ref44} are PADE-I and PADE-II \cite{ref45}. These are some well known and
most used dark energy parameterization in this series. We used these five well-known parameterization of dark energy, 
namely CPL, JBP, BA, PADE-I and PADE-II in our model and also find the swampland correspondence with our related model.\\

The paper is structured as follows: For Sec. $2$, in the context of $f(R, T)$ gravity, we have developed the basic field equations 
and derived the related geometric parameters. Section $3$ describes the dynamics of the model. Stability and physical acceptability 
of the solution are discussed in Sec. $4$. Parameterization of dark energy is addressed in Sec. $5$. Section $6$ shows the 
correspondence of swampland criteria of dark energy. In the last section $7$, the description and conclusion 
are given.


\section{Metric and Field Equations}
Right now  in this section we  talk about  the formalism created in an insignificantly coupling theory of $f(R,T)$ to investigate 
certain models. We consider a Bianchi $VI_{h}$ space time
\begin{equation}
\label{1}
ds^{2}=dt^{2}-A^{2}dx^{2}-B^{2}{e^{2x}} dy^{2}-C^{2} e^{2hx} dz^{2},
\end{equation}

where  $A(t)$, $B(t)$ and $C(t)$ are functions of the cosmic time $t$. The space-time exponent $h$ is taken the value as $-1,0,1$. Here, we 
considered the  $h=-1 $ because of the significance of the metric that envisages an isolated universe with invalid total energy and momentum
 \cite{ref46}\cite{ref47}.

\begin{equation}
\label{2}
T_{ij}=(p+\rho)u_{i}u_{j}-\rho_{B}x_{i}x_{j}-pg_{ij}.
\end{equation}

Here  $u^{i} x_{i}=0 $ and $x^{i}x_{i}=-u^{i}u_{i}=-1$. The four velocity vector
of the fluid in a co-moving model is $u^{i}=\delta_{0}^{i}$. $x^{i}$ is anisotropic fluid direction and orthogonal to $u^{i}$. Such as the
Perfect fluid and anisotropic fluid ($\rho_{B}$), with energy density $(\rho)$.  Harko {\it et al.} \cite{ref10}
proposed the principle of $f(R, T)$ theory on the interaction of matter and geometry.The four-dimensional Einstein-Hilbert movement is composed as:

\begin{equation}
\label{3}
S=\frac{1}{16\pi}\int d^4 x\sqrt{-g}f(R,T) + \int d^4 x\sqrt{-g})L_{m},
\end{equation}

where $f(R, T)$ is a function of $T(=g_{ij} T^{ij})$ and Ricci scalar $R$ in the operation, $T^{ij}$ is the energy- momentum tensor. 
It is possible to take Lagrangian  $L_{m}$  as either $L_{m}= -p$ or as $L_{m}=\rho$. We consider a minimal coupling of geometry 
and curvature for the modified gravity model, assuming $f(R, T)=f(R)+f(T)$. We can write the field equations as \cite{ref25}-\cite{ref48} 
after our earlier works.

\begin{equation}
\label{4}
f_{R}(R) R_{ij}-\dfrac{1}{2}f(R)g_{i,j}+ (g_{ij} \square-\nabla_{i}\nabla_{j})+f_{R}(R)=[8\pi+f_{T}(T)] T_{ij}+[pf_{T}(T)+\dfrac{1}{2}f(T)]g_{ij},
\end{equation}

where the corresponding partial differentiations are $f_{ R}=\dfrac{\partial f(R)}{\partial R}$ and $f_{T}=\dfrac{\partial f(T)}{\partial T}$.
The field equations \cite{ref48} derive from a specific choice of $f(R, T)$=$\lambda(R+T)$
\begin{equation}
\label{5}
G_{ij}= \left(\dfrac{8\pi+\lambda}{\lambda}\right)T_{ij}+\Delta(T)g_{ij},
\end{equation}

where $\lambda$ is a non-zero scaling variable in GR. $G_{ij}= R_{ij}-\dfrac{1}{2}Rg_{ij}$ is the Einstein tensor. A time-dependent
effective cosmological constant can be defined with the variable $\Lambda(T) = p+\frac{1}{2}T$ occurring in the field Eq. (\ref{5}).
Here $\Lambda(T)$ depends on the substance of the matter content and helps to accelerate. Nevertheless,  Eq. (\ref{5}) has the same 
mathematical form of GR with a time varying constant, due to the non-disappearing quantity of $\lambda$ it can not be reduced to GR. 
Eq. (\ref{5}), however, is a rescaled generalization of GR equations. The field Eq. (\ref{5}) can be written specifically for Bianchi 
form $VI_{h}$ space-time in modified gravity as:

\begin{equation}
\label{6}
\dot H_{y}+H_{y}^2+\dot H_{z}+H_{z}^2++H_{y}H_{z}+\frac{1}{A^{2}}=-\left(\frac{16\pi+3\lambda}{2\lambda}\right)
\left(p-\rho_{B}\right)+\dfrac{\rho}{2},
\end{equation}

\begin{equation}
\label{7}
\dot H_{x}+H_{x}^2+\dot H_{z}+H_{z}^2+H_{x}H_{z}-\frac{1}{A^{2}}=-\left(\frac{16\pi+3\lambda}{2\lambda}\right) p + 
\left(\dfrac{\rho_{B}+\rho}{2}\right),
\end{equation}

\begin{equation}
\label{8}
\dot H_{x}+H_{x}^2+\dot H_{y}+H_{y}^2++H_{x}H_{y}-\frac{1}{A^{2}}=-\left(\frac{16\pi+3\lambda}{2\lambda}\right) p+
\left(\dfrac{\rho_{B}+\rho}{2}\right),
\end{equation}

\begin{equation}
\label{9}
H_{x}H_{y}+H_{y}H_{z}+H_{x}H_{z}-\frac{1}{A^{2}}=\left(\frac{16\pi+3\lambda}{2\lambda}\right) \rho -\left(\dfrac{p-\rho_{B}}{2}\right),
\end{equation}
\begin{equation}
\label{10}
H_{y}-H_{z}=0.
\end{equation}

Here $\lambda$ is the non zero scale factor. Directional Hubble parameters for the anisotropic model are $H_{x}=
\frac{\dot A}{A}$, $H_{y}=\frac{\dot B}{B}$ and  $H_{z}=\frac{\dot C}{C}$. In view of Eq. (\ref{10}) we have  $H_{y}=H_{z}$. 
Assuming  $H_{x}=m H_{z}$ for $m\neq1$, $H$ is the mean Hubble parameter can be defined as:

\begin{equation}
\label{11}
H=\frac{\dot a}{a}=\frac{1}{3}(H_{x}+H_{y}+H_{z})=\frac{1}{3}\left(\frac{\dot A}{A}+\frac{\dot B}{B}+\frac{\dot C}{C}\right).
\end{equation}

By using Eq. (\ref{10}), we get 

\begin{equation}
\label{12}
H=\frac{\dot a}{a}=\frac{1}{3}(H_{x}+2 H_{y})=\left(\frac{m+2}{3}\right)H_{z}.
\end{equation}
Here $q$ is as a linear function of $H$ then

\begin{equation}
\label{13}
q=-\frac{a \ddot a}{\dot a^{2}} = \frac{\beta}{\sqrt{2 \beta t+k}} -1.
\end{equation}

In addition, as indicated by SNe Ia's despise resent observation, the present universe is extending and the estimation of DP 
is in the scope of $-1<q<0 $ sooner or later. This is the reason our model is perfect with resent observation, in every one of 
the three circumstances.
\begin{equation}
\label{14}
q_{0}=-1+\beta H_{0} = -1+\frac{\beta}{\sqrt{2\beta t_{0}+k}}.
\end{equation}

Since the present estimation  of declaration parameter can be taken as here $k$ and $\beta$ are positive constants. From Eq. (\ref{13}) 
we observed that $q>0$ for $\frac{\beta}{\sqrt{2\beta t+k}}<1$ and $q<0$ for$\frac{\beta}{\sqrt{2\beta t+k}}>1$. so we can choose a 
value of $\beta$ and $k$.\\
 
Integrating Eq. (\ref{13}) then we get a scale factor (a) which is depends on time.

\begin{equation}
\label{15}
a(t)=e^{\frac{1}{\beta}\sqrt{2 \beta t + k}}.
\end{equation}

We can express the directional Hubble parameters as $H_{x}=e^{\frac{3m}{\beta(m+2)}\sqrt{2 \beta t + k}}$ and  
$H_{y}=H_{z}=e^{\frac{3}{\beta(m+2)}\sqrt{2 \beta t + k}}$, 
$A=e^{\frac{3m}{\beta(m+2)}\sqrt{2 \beta t + k}}$, $B=C=e^{\frac{3}{\beta(m+2)}\sqrt{2 \beta t + k}}$,\\

so that the metric (\ref{1}) can be written as 

\begin{equation}
\label{16}
ds^{2}=dt^{2}-(e^{\frac{6m}{\beta(m+2)}\sqrt{2 \beta t + k}})dx^{2}-(e^{\frac{6}{\beta(m+2)}\sqrt{2 \beta t + k}})e^{2x}dy^{2} - 
(e^{\frac{6}{\beta(m+2)}\sqrt{2 \beta t + k}})e^{2hx}dz^{2}.
\end{equation}
The spatial volume $V$ is given as

\begin{equation}
\label{17}
V=(AB^{2}) = a^3(t).
\end{equation}

The Hubble parameter is  defined as:
\begin{equation}
\label{18}
H=\frac{1}{3}(H_{x}+2H_{y})=\dfrac{1}{\sqrt{2\beta t+k}}.
\end{equation}
The scalar expansion in the universe is
\begin{equation}
\label{19}
\theta=3H=(H_{x}+2H_{y})=\dfrac{3}{\sqrt{2\beta t+k}}.
\end{equation}
The shear scalar is
\begin{equation}
\label{20}
\sigma^{2}=3\left(\dfrac{m-1}{m+2}\right)^2\dfrac{1}{\sqrt{2\beta t+k}}.
\end{equation}
The cosmic jerk parameter $j$ in cosmology is characterized as the third derivative of the scale factor concerning the astronomical 
time is given as

\begin{equation}
\label{21}
j=\frac{1}{H^{3}}\left(\frac{\dddot a}{a}\right)=\left(q+2q^2-2\frac{\dot q}{H}\right).
\end{equation}

 In cosmology the jerk parameter is used to describe the models close to $\Lambda CDM$. It is acknowledged that for models with 
 negative estimation of the deceleration parameter and positive estimation of the jerk parameter, the transition of the universe 
 from decelerated stage to accelerated stage happens. The $\Lambda CDM$ jerk parameter has consistent with $j=1$. Right now, we get
\begin{equation}
\label{22}
j=1-\frac{3\beta}{\sqrt{2\beta t+k}}+\left(\frac{3\beta^{2}}{{2\beta t+k}}\right).
\end{equation}

Here we take three observational data  and find the value of ($\beta$, $k$) to be the best fit with the latest observations and 
considered for drawn all figures.

\section{ Dynamics of the Model}

In this section we can obtain physical and kinematic parameters  from the Eq. (\ref{6})-(\ref{11}).

\begin{equation}
\label{23}
\rho=\frac{6}{1-4\alpha^{2}}\left[\frac{2}{m+2}\left(\dot H+H^2\right)+\frac{(5-2m)-6\alpha (2m+1)}{(m+2)^2}H^2\right]+
\frac{2 a^{-\frac{6m}{m+2}}}{1-2\alpha}
\end{equation}

\begin{equation}
\label{24}
\rho_{B}=\frac{6}{1-2\alpha}\left[\frac{m-1}{m+2}\left(\dot H+3H^2\right)\right]-\frac{4 a^{-\frac{6m}{m+2}}}{1-2\alpha}
\end{equation}

\begin{equation}
\label{25}
p=\frac{6}{1-4\alpha^{2}}\left[\frac{(m-1)+2 \alpha (m+1)}{m+2}\left(\dot H+H^2\right)+\frac{(2m^2-4m-7)+2\alpha (2m^2+1)}{(m+2)^2}
H^2\right]-\frac{2 a^{-\frac{6m}{m+2}}}{1-2\alpha}
\end{equation}

The effective cosmological constant $\Lambda$ and EoS parameter ( $\omega$) are other complex features of the model. Such parameters 
are calculated using the scale function.

\begin{equation}
\label{26}
\omega=\left[ \frac{(1+2\alpha)\{3(m^2+3m+2)(\dot H+H^2)+(6m^2-18m-6)H^2 \}}{6(m+2)(\dot H+H^2)-3(2m-5)H^2+(m+2)^2 a^{-\frac{6m}{m+2}}
-2\alpha] \left(9(2m+1)H^2-(m+2)^2 a^{-\frac{6m}{m+2}}\right)}\right] - 1
\end{equation}

\begin{equation}
\label{27}
\Lambda=\frac{6}{(1+2\alpha)(m+2)}\left[\dot H+3H^2\right].
\end{equation}

Here we use $\alpha= (16\pi+3\lambda)/2\lambda$. All the physical parameters are expressed in terms of Hubble parameter $(H)$ in the 
above equations. 

\begin{figure}[H]
	\centering
	(a)\includegraphics[width=7cm,height=7cm,angle=0]{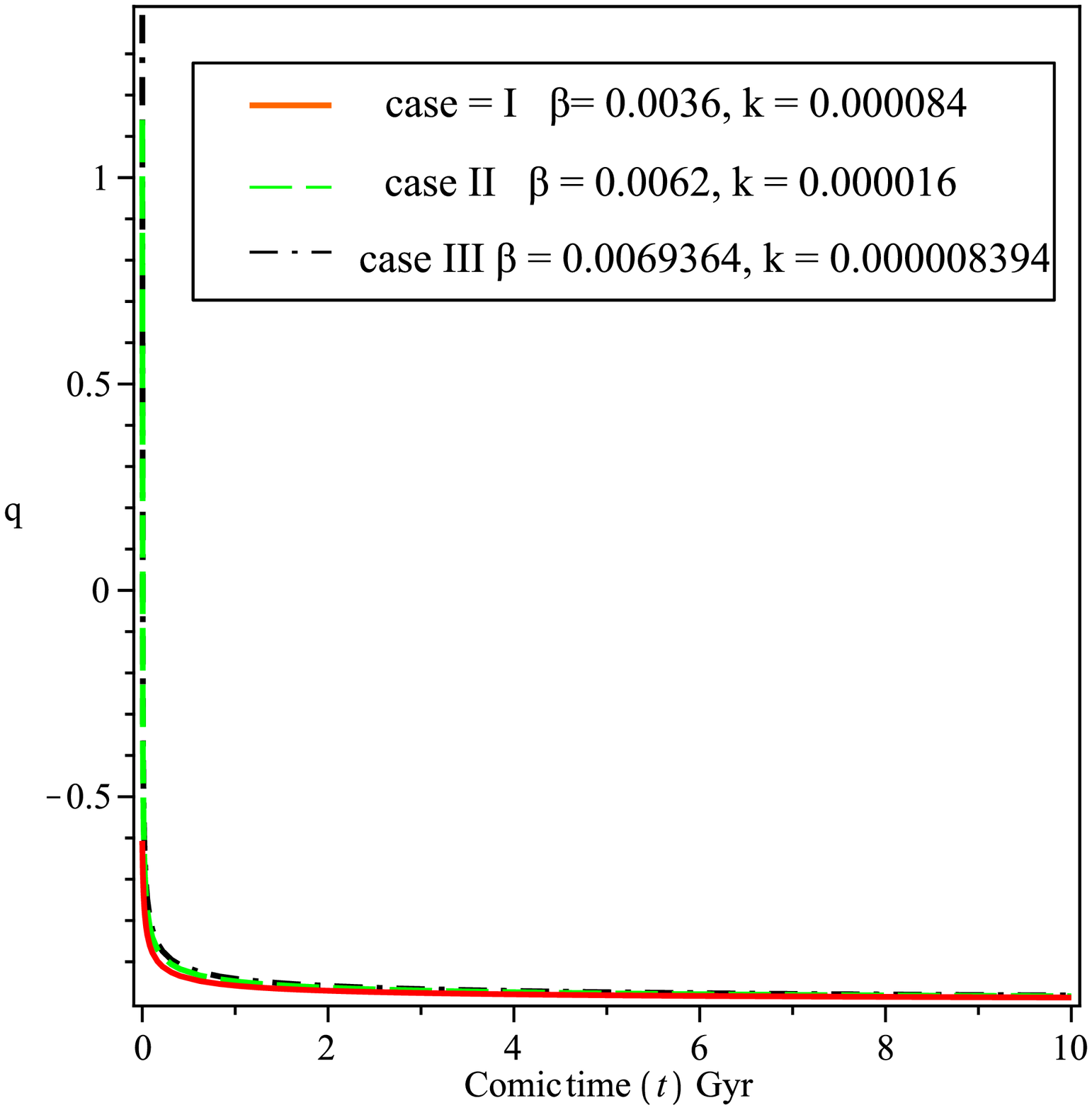}
	(b)\includegraphics[width=7cm,height=7cm,angle=0]{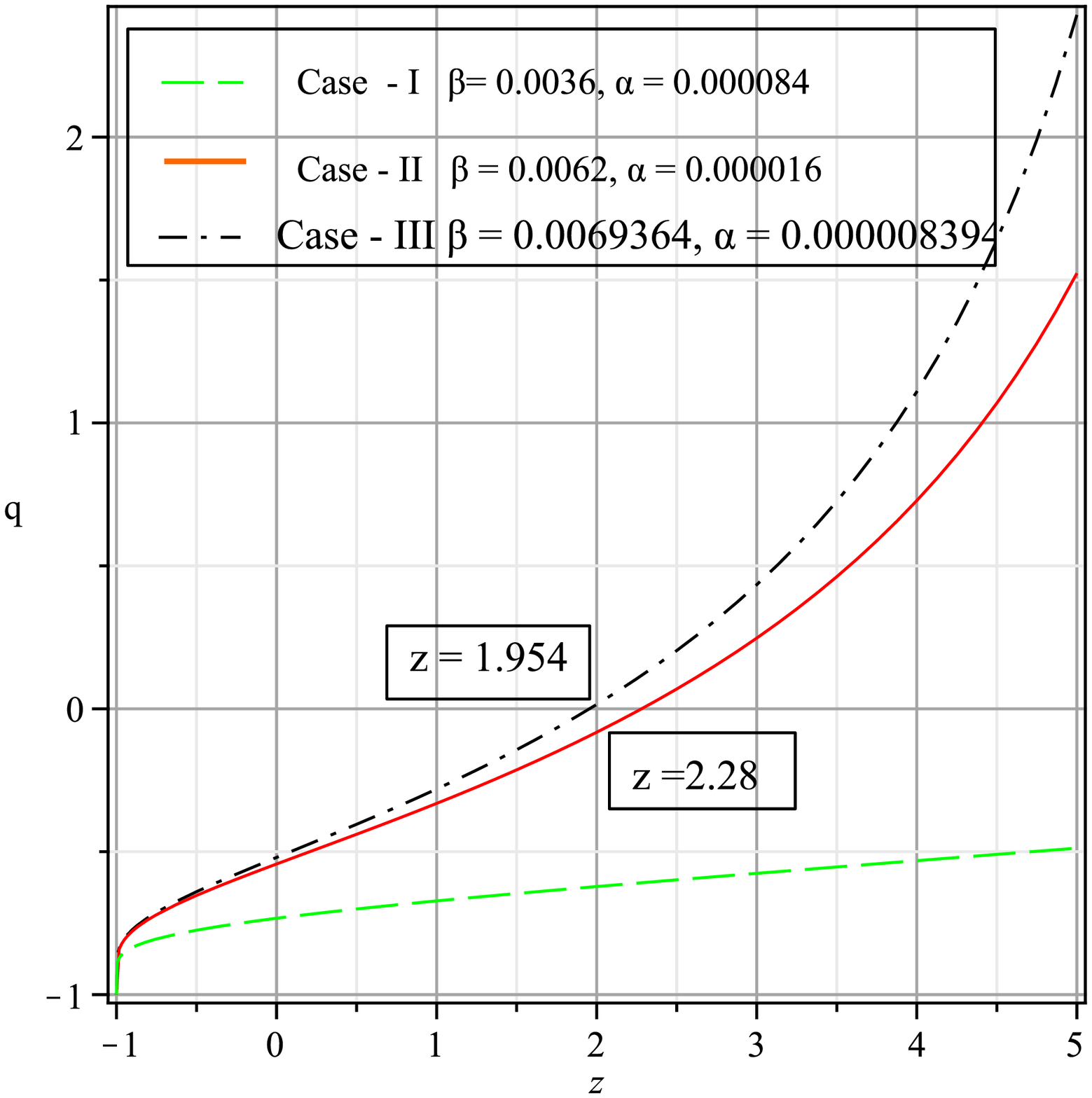}
	\caption{(a) Plot of DP ($q$) versus $t$ 
		 (b) Plot of DP ($q$) versus redshift $z$} 
\end{figure}

	\begin{table}[H]
	\caption{\small According to three observed data's, the corresponding values of
		 of DP $(q_{0})$  and Hubble parameter ($H_{0}$), we find the values of $\beta$ and k .}
	\begin{center}
		\begin{tabular}{|c|c|c|c|c|c|c |}
			\hline
		\tiny Cases & \tiny Data  &\tiny	$q_{0}$  &	\tiny  $H_{0}$ & \tiny $\beta$ & \tiny $k$ & \tiny Reference \\
			\hline
			
	\tiny Case-I & \tiny Supernova type Ia Union  &	 \tiny	-0.73	& \tiny 73.8	 &\tiny 0.0036  & \tiny 0.000084 & \tiny \cite{ref49}\\ 
			\hline	
			
\tiny Case-II & \tiny  BAO and CMB  &\tiny $-0.54$ & \tiny $73.8$ &\tiny 0.0062& \tiny 0.000016 &\tiny \cite{ref50} \\ 
			\hline

\tiny Case-III & \tiny OHD+JLA  & \tiny	-0.52	& \tiny 69.2	 &\tiny 0.000069364& \tiny 0.000008394 & \tiny \cite{ref51}-\cite{ref52} \\
			\hline

\end{tabular}
\end{center}
\end{table}


Figure $1(a)$ shows the assortment of deceleration parameter $(q)$ with astronomical time $(t)$ as indicated by Eq. (\ref{13}). We just observe 
our design model is an acceleration stage ($q<0$) for $k= 0.000084 $ and $\beta=0.0036 $ (case I). In (case-II)   $k =0.000016 $ and
$\beta= 0.0062 $ our model depicts the phase transition from +ve to -ve  deceleration parameter$(q)$.
This shows that our 
model is changing from $(q>0)$ deceleration to $(q<0)$ acceleration. $t_{c} = \dfrac{\beta^{2}-c}{2\beta}$ is the critical time at which 
the phase transaction took place. Similarly, \cite{ref51, ref52} used the value of the joint dataset (OHD+JLA) in which  $k=0.0069364$ and 
$\beta=0.000008394 $ we find decelerating-accelerating phase transition (case III) (Table-1). \\

The average scale factor $a(t)$ as far as redshift $z$ is given by $a(t)=\frac{a_0}{1+z}$, From Eq. (\ref{15}), we get  
$\frac{\sqrt{2\beta t + k}}{\beta}= ln(a)$, then $ln(a)= ln(a_0)-ln(1+z)$ . Substituting the above in Eq. (\ref{13}), we shows 
 $q$-parametrization given by

\begin{equation}
\label{28}
q(z)=-1+\frac{1}{ln(a_0)-ln(1+z)},
\end{equation} 
where $a_0$ is present value of scale factor, for three cases of $(\beta, k)$  $(\beta=0.0036, k=0.000084, \beta=0.0062, k=0.000016, 
\beta=0.00006, k=0.00000 )$ for plot the graphs. In the figure 1(b) we have discovered that $q(z)$ expansion is a smooth progress from a 
decelerated stage to accelerated period of development and $q\to -1$ as $ z \to -1$. As of late \cite{ref53}-\cite{ref54} have 
discovered the change redshift from decelerating to accelerated in modified gravity cosmology. SNe type Ia dataset has given 
the progress from past deceleration to ongoing increasing speed at $\Lambda CDM$. More recently, in 2004 (HZSNS) 
team have identified $ z_t = 0.46 \pm 0.13$ at $(1\sigma )$ c.l. \cite{ref2} which is again improved to 
$z_t  = 0.43 \pm 0.07$ at $(1\sigma )$ c.l. \cite{ref7}. SNLS \cite{ref55}, and recently 
complied by \cite{ref56}, provide a progress redshift $z_t  + 0.6(1\sigma )$ in improved concurrence with the flat $\Lambda CDM$.\\

In addition, the $q(z)$ reconstruction is performed by the Joined
$(SNIa + CC + H_0 )$, which have acquired the change redshift 
$ z_t  = 0.69^ {+0.9}_{-0.6} $, $0.65^ {+0.10}_{-0.07}$ and $0.61^ {+0.12}_{-0.85} $ inside $(1\sigma)$ \cite{ref57}. which are seen 
as well predictable with past outcomes \cite{ref58}-\cite{ref62} including the $\Lambda CDM$ expectation $z_t\approx 0.7$. Another 
constraint of change redshift is $0.60\leq z_t \leq 1.18$  ($2\sigma$ joint examination) \cite{ref63}. From the $H(z)$ data we find 
evidence as $z_{t}=0.720\pm 0.14$ for redshift. which is in acceptable 
concurrence with the \cite{ref64} assurance of $zt= 0.72\pm 0.05$ as well as \cite{ref65} assurance of $z_t= 0.82\pm 0.08$ at 
$1\sigma$ error. This is again improved as $z_t= 0.72\pm 0.14 $ at $68\%$ c.l. which is in 
acceptable understanding. From the combination of $H(z)$ and SNIa datasets, we note that the transition from deceleration to acceleration 
in the BI expansion process takes place at a redshift of $z_t = 0.57\pm 0.0037$ which is in acceptable concurrence with the outcomes 
acquired \cite{ref66,ref67}. Therefore, with reference to resent observation of SNe Ia, the present 
universe is expanding and at some stage the value of DP is within the range of $-1<q<0$. So, we observe that in all three cases, our model is 
aligned with resent findings.\\

The transition redshift for our derived models for two cases (iii) $\beta = 0.000008394$, $k = 0.0069364$ and 
(ii) $\beta = 0.0062$, $k = 0.000016$ are found to be $z_{t} = 1.954$ and $z_{t} = 2.28$ respectively (Fig. $1(b)$) which are in good 
agreement with observational values. 

\begin{figure}[H]
	(a)\includegraphics[width=8cm,height=8cm,angle=0]{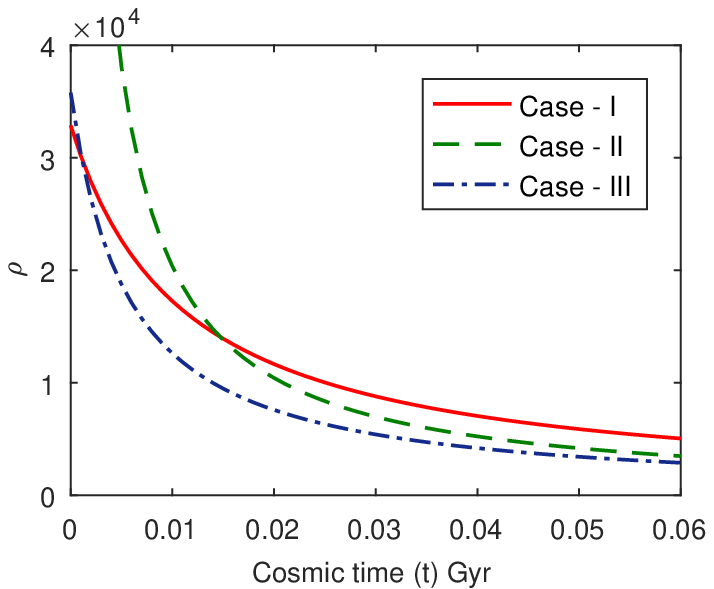} 
	(b)\includegraphics[width=9cm,height=8cm,angle=0]{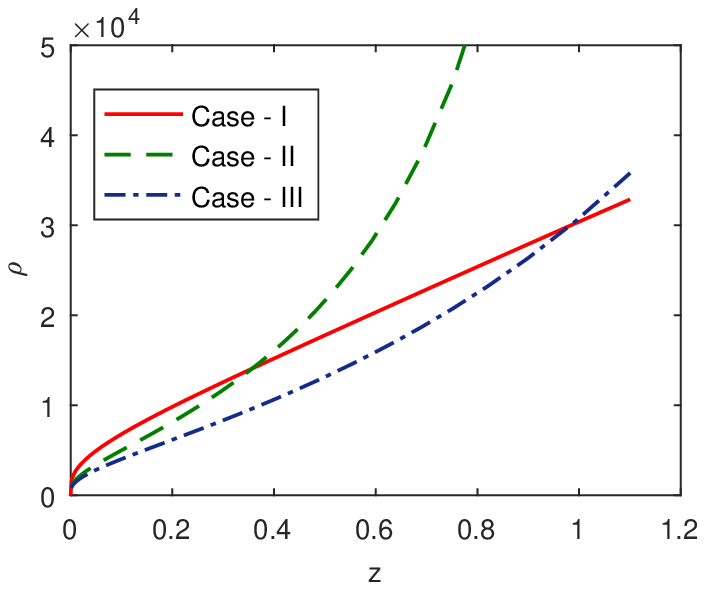}
	\caption {Plot of energy density $\rho$ versus $t$ and redshift. Here $\lambda = 1$, $\alpha = 26.6$, $ m= 0.004$. }
	\end{figure}

Figure $2(a,b)$ which is related to the Eq. (\ref{28}), depict the energy density $\rho$ with  time $t$ and redshift $z$ for three cases. It is 
seen that $\rho $ remains positive during the infinite development. Here we additionally noticed that $t\to 0$, $\rho\to\infty$ showing 
the big-bang situation. Our model is likewise steady with ongoing perceptions. We conclude that in fig.2(a) $\&$ fig.2(b) the 
$\rho$ is a positive diminishing function of time and increasing function of $z$. It approaches  to zero as $t\to\infty$. Which is 
consistent with the well establishment scenario. 

\begin{figure}[H]
	\centering
	(a)\includegraphics[width=7cm,height=7cm,angle=0]{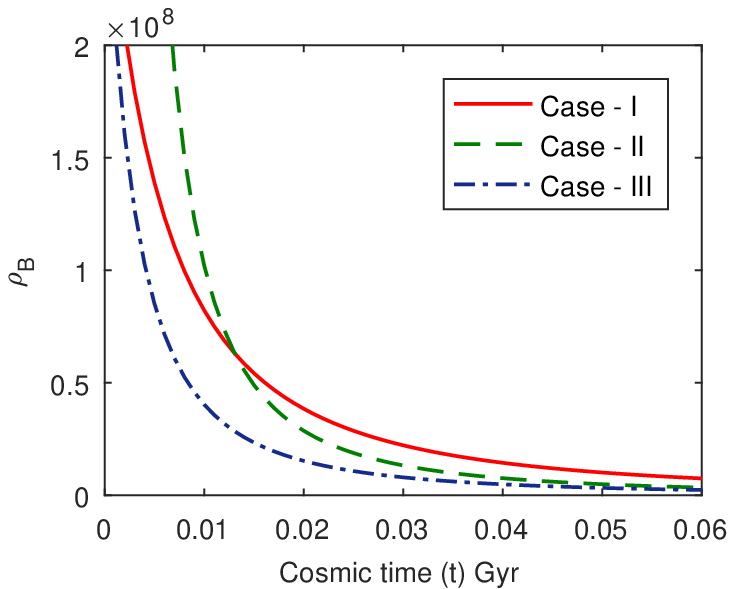}
	(b)\includegraphics[width=7cm,height=7cm,angle=0]{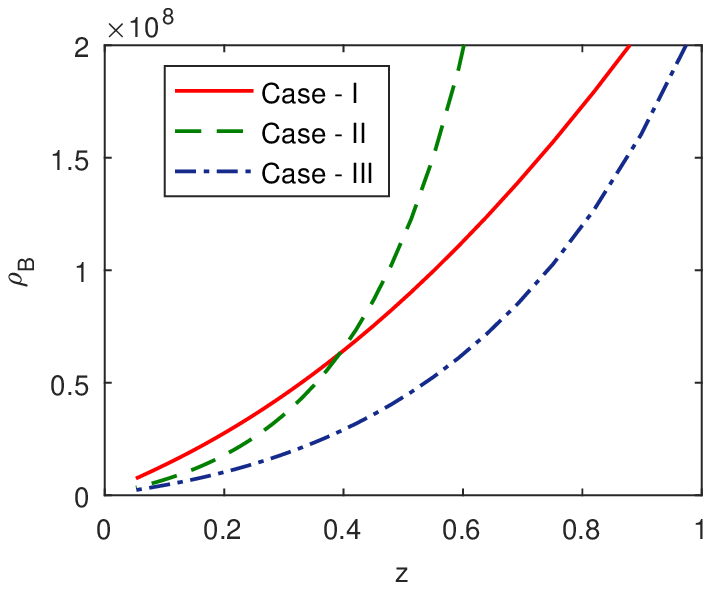}
	\caption{Plot of anisotropic fluid energy density  $(\rho_{B})$ versus $t$ and redshift. 
	Here $\lambda = 1$, $\alpha = 26.6$, $ m= 0.004$. } 
\end{figure}
Figure $3$ displays anisotropic fluid energy density in terms of cosmic time $t$ and redshift $z$. Eq. (\ref{29}) corresponding to anisotropic fluid 
energy density $(\rho_{B})$ is a declining time function and remains positive during the cosmic assessment. First, we can see in the figure 3(a), 
it fell sharply, then gradually, and at the present epoch, it approached  to a small positive value. Here $(\rho_{B})$ tends to 0 as $t\to\infty$. 
But in Fgure 3(b)  anisotropic fluid energy density increases with redshift
because density decreases imply the volume increase, the expansion of the universe. In all three cases that the anisotropic density of the dark
matter decreases with time and tends to zero at late times. 
\begin{figure}[H]
	\centering
	(a)\includegraphics[width=7cm,height=7cm,angle=0]{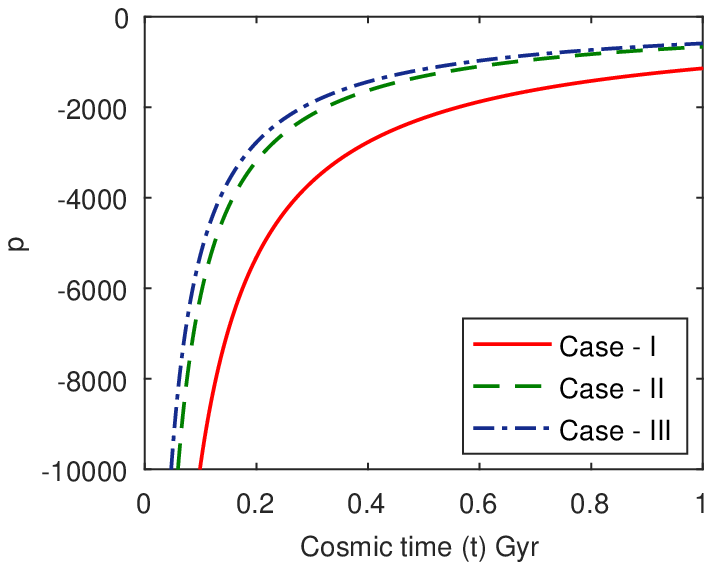}
	(b)\includegraphics[width=7cm,height=7cm,angle=0]{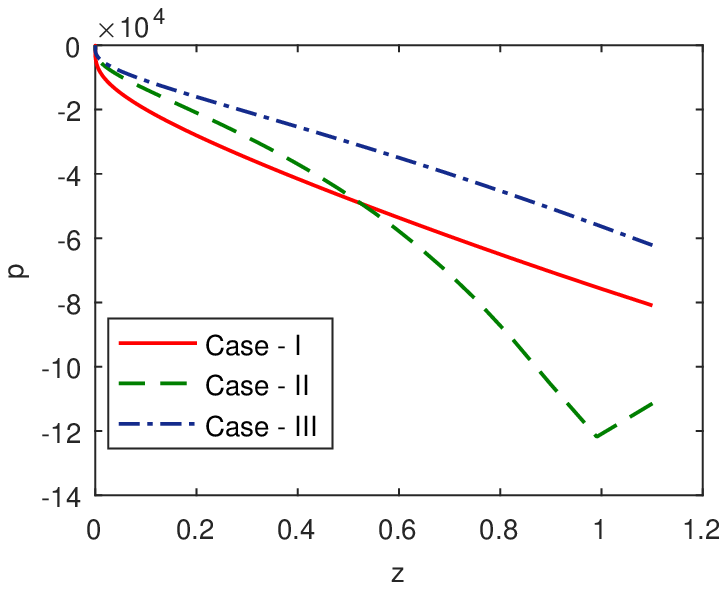}
	\caption{Plot of pressure $p$ versus $t$ and redshift. Here $\lambda = 1$, $\alpha = 26.6$, $ m= 0.004$.} 
\end{figure}

Figure $4$ represents the expansion for fluid pressure $p$ for the model and corresponding to (\ref{25}). We observed that for all three 
cases pressure is negative increasing function of time $t$ and redshift $z$. From the figures $4(a)$ and $4(b)$ we observed for the homogeneous and 
isotropic model, pressure is consistently negative and approaches zero at late time.


\begin{figure}[H]
	\centering
	(a)\includegraphics[width=7cm,height=7cm,angle=0]{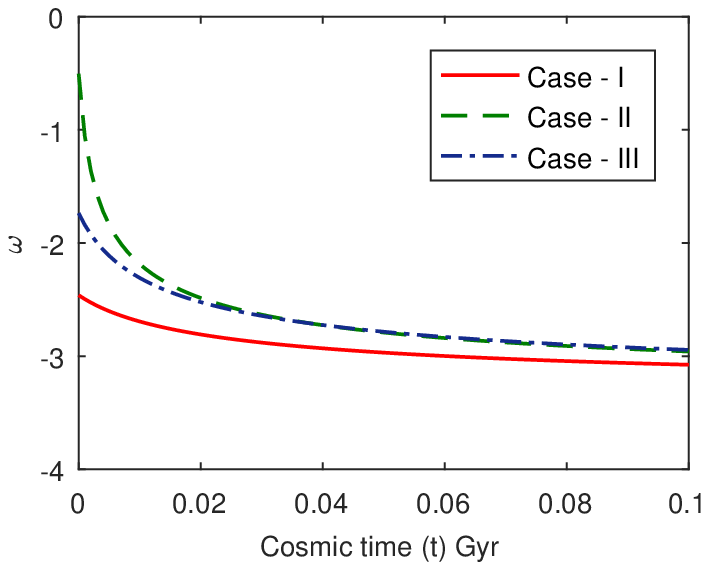}
	(b)	\includegraphics[width=7cm,height=7cm,angle=0]{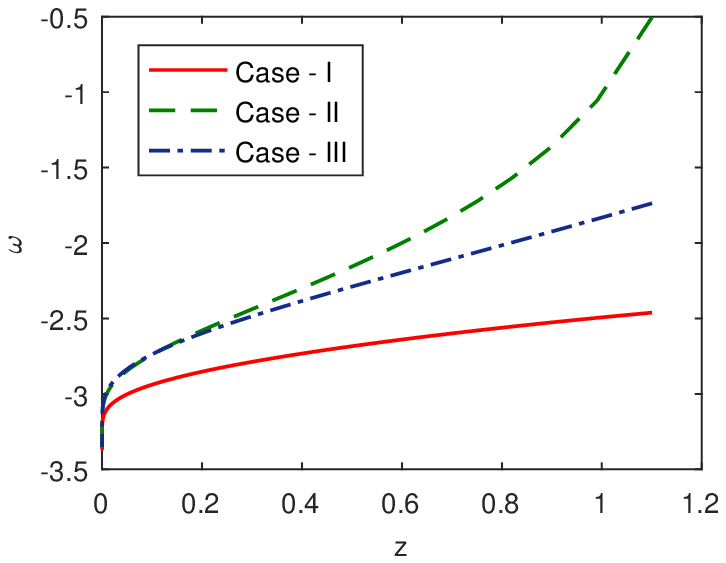}
	\caption{ Plot  of EoS parameter $\omega$ versus $t$ and redshift. Here $\lambda = 1$, $\alpha = 26.6$, $ m= 0.004$.}
\end{figure}
Figs. 5(a) and  5(b) depict the behavior of EoS parameter $\omega$ with respect to time $t$ and redshift $z$ respectively. The EoS parameter 
may lies in phantom region $(\omega <-1)$. Both figures represent the same scenario of the universe for all cases.
 It is clearly shows in fig. 5(a) that first equation of state parameter 
decreased sharply and approach to a small negative values at the present epoch. The fig. 5(b) shows that the EoS parameter is a 
increasing function of $z$, the rapidity of its growth at the early stage deponds on the types of the universe.

\begin{figure}[H]
	\centering
	(a)\includegraphics[width=7cm,height=7cm,angle=0]{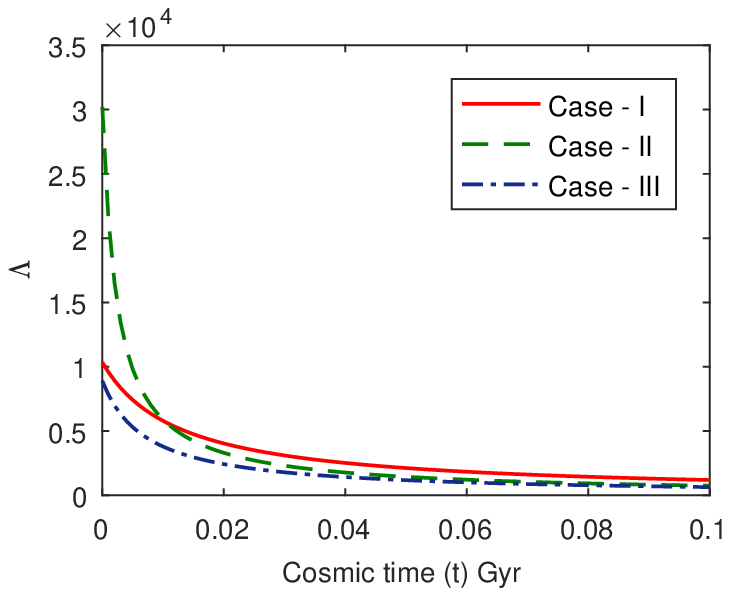}
	(b)	\includegraphics[width=7cm,height=7cm,angle=0]{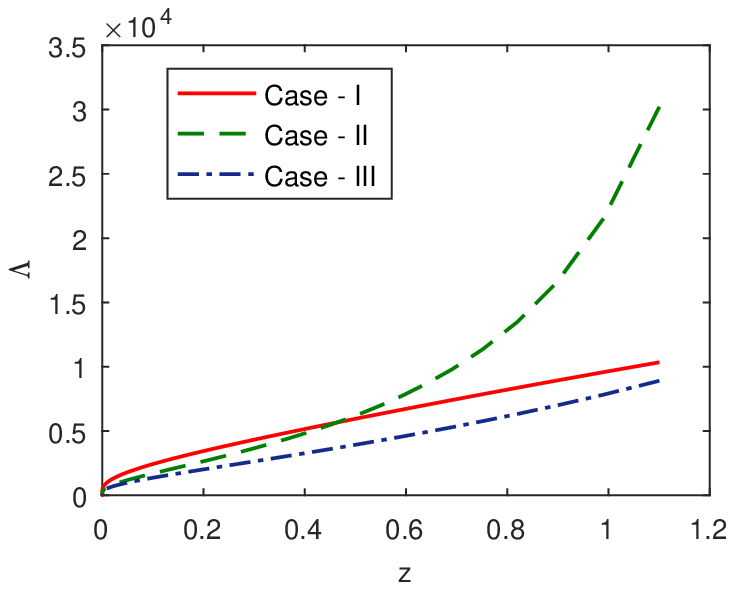}
	\caption{Plot of cosmological constant $\Lambda$ versus $t$ and redshift. Here $\lambda = 1$, $\alpha = 26.6$, $ m= 0.004$. } 
\end{figure}

Figure 6(a) and 6(b) depicts the cosmological constant with respect to time and redshift $z$. In all cases $\Lambda$ is decreasing 
function of time and increasing function of redshift $z$. We expect that in the universe,  the positive value of $\Lambda $ the expansion 
will tends to accelerate. In fig. 6(a) we observed that the $\Lambda$ decreases more sharply as time increases in empty universe as 
compared to radiating dominated and stiff fluid universe. Here fig.6(b) shows that $\Lambda$ increases with high redshift. A positive 
cosmological constant resists attractive gravity of matter due to its negative pressure.

\section { Physical Acceptability and Stability  of the Outcomes}
Now, we shall test by means of some recently used diagnostic tools, whether our models are stable or not ? 

\subsection{Energy Condition}
In this subsection, for our derived model, we are checking the energy conditions. The three energy conditions (NES)$ (\rho+p\geq0)$, 
(SEC)$ (\rho+3p)\geq 0$ and  (DEC)$ (\rho-p)\geq 0 $ are given. In Figs. $7(a)$, $7(b)$ and $7(c)$, we have plotted the energy conditions 
with respect to cosmoc time for all three cases I, II \& III. From these figures we observe that energy conditions violet in all 
cases which is aspected in the case of DE.


\begin{figure}[H]
	\centering
(a)	\includegraphics[width=5cm,height=5cm,angle=0]{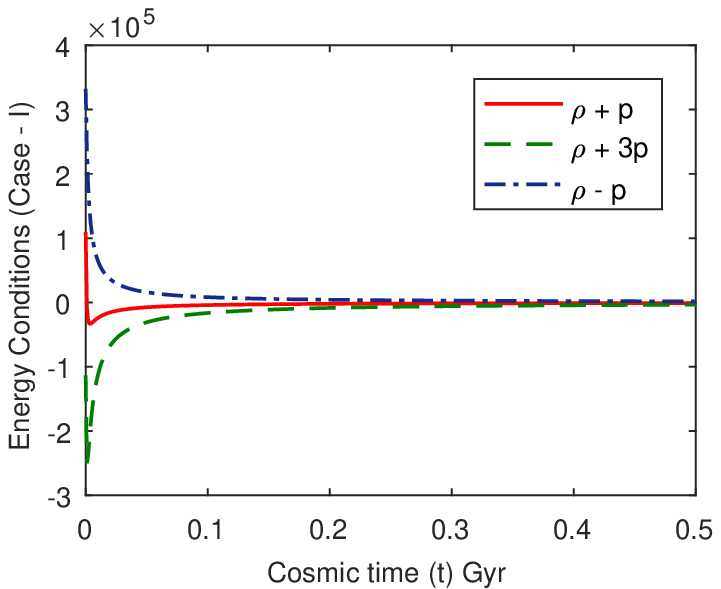}
(b)	\includegraphics[width=5cm,height=5cm,angle=0]{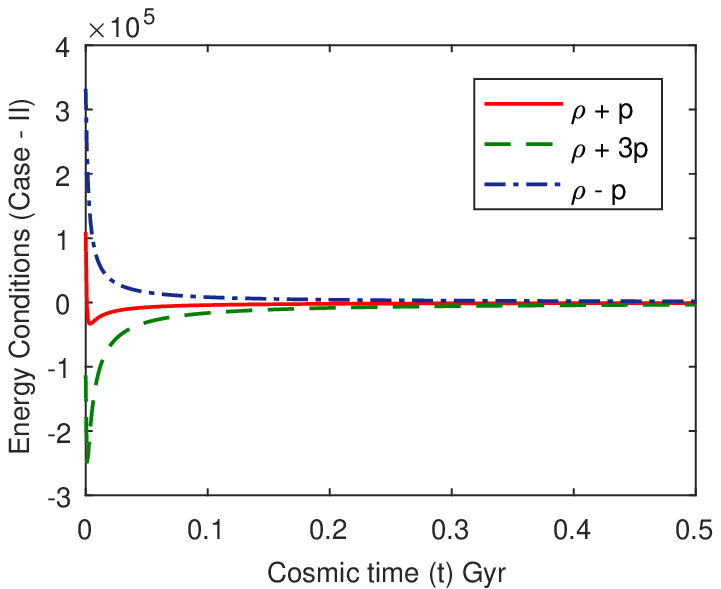}
(c) \includegraphics[width=5cm,height=5cm,angle=0]{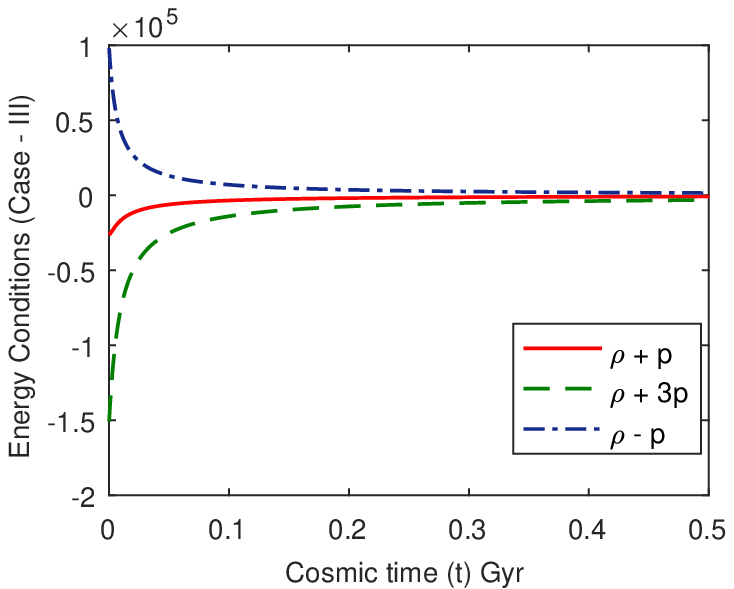}
	\caption {Plot of energy conditions (NEC,SEC,DEC) with $t$. Here $\lambda = 1$, $\alpha = 26.6$, $ m= 0.004$.}
\end{figure}

\subsection{Velocity of Sound}

Numerous authors \cite{ref68}-\cite{ref70} have researched the requirements on sound speed of dynamic DE models with time-changing 
EoS ($\omega$) and presumed that imperative on the sound speed of DE is exceptionally weak. DE parameters just as other cosmological 
parameters are independent of the compelling sound speed of DE, there is no limitation from present astronomical information on the effective 
sound speed. The system is unstable if squared speed  of sound $v_s^{2}<0 $. It is necessary that the $v_s^{2}$ sound speed should be 
less than the velocity 
of light $c$. As we work with unit speed of light in gravitational units, i.e. sound velocity exists within the range 
$0 \leq v_s^{2} = ( \frac{dp}{d\rho}) \leq 1$. We get the velocity of  sound as, \\

\[v_s^{2}=\frac{6}{1-4\alpha^2}\left[{\frac{(m-1)+2\alpha(m+1)}{m+2}}\left(\frac{-2\beta}{(2\beta t+k)^2}+\frac{3 (\beta)^2}
{(2\beta t+k)^\frac{5}{2}}\right)-
2\beta\frac{(2m^2-4m-7)+2\alpha(2m^2+1)}{(m+2)^2(2\beta t+k)^2}\right]\]

\[+12m\frac{4 e^\frac{-6m\sqrt{2\beta t+k}}{\beta (m+2)}}{(m+2)\sqrt{2\beta t+k}(1-2\alpha)} \times
\] 
\[\frac{1}{ \frac{6}{1-4\alpha^2} \left[\frac{2}{m+2}\left(\frac{-2\beta}{(2\beta t+k)^2}+\frac{3 (\beta)^2}{(2\beta t+k)^\frac{5}{2}}\right)-
	2\beta\frac{(-2m+5)+6\alpha(2m+1)}{(m+2)^2(2\beta t+k)^2}\right]
}
\]
\begin{equation}
\label{29}
-12m\frac{4 e^\frac{-6m\sqrt{2\beta t+k}}{\beta (m+2)}}{(m+2)\sqrt{2\beta t+k}(1-2\alpha)} 
\end{equation}


\begin{figure}[H]
	\centering
	\includegraphics[width=7cm,height=7cm,angle=0]{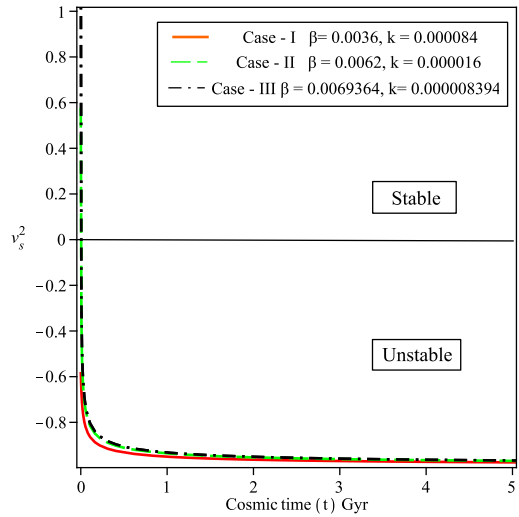}
	\caption{Plot of velocity of sound versus $t$. Here $\lambda = 1$, $\alpha = 26.6$, $ m= 0.004$. } 
\end{figure}
Figure 8  dipicts that a positive value of squared speed of sound $v_s^{2}>0 $ represents a stable model. Here we notice that in case I, our 
model is unstable as $v_s^{2} < 0$ through whole evolution of the Universe but in case II and case III, our model is stable in early phase 
of the Universe whereas unstable in present phase.

\section{Dark Energy Parameterization}

Given a large number of scalar field models with a range of potentialities, testing all individual models is always difficult. 
Alternatively, we often use a parameterization of the evolution of dark energy that broadly describes a large number of DE
models in the scalar region. In parameterizing  $(\omega= p/\rho)$ dark energy equation of state is the most common practice. So, 
as to evaluate the dynamical parts of the model, we have plotted the EoS parameter $\omega$ as an element of redshift and the 
literature\cite{ref71}-\cite{ref82} also contains a large number of parameterizations for $\omega $. The conduct of the EoS parameter 
 of the model has been contrasted and that of some notable EoS parameterization like CPL \cite{ref39}-\cite{ref42}, JBP[43], BA \cite{ref44}, 
 PADE-I and PADE-II \cite{ref45} for our measurements of the evidence together with the regular $\Lambda CDM$ and the constant dark energy 
 equation of state model. 
 
 Now, in this segment we consider the five well known DE parameterization. The first is the Chevallier-Polarski-Linder model (CPL), where $\omega_{0}$ is the 
 present EoS value and its overall time evolution  $\omega_{a}$ is the CPL model written as follows.
\begin{equation}
\label{30}
\omega=\omega_{0}+\omega_{a}\frac{z}{(1+z)}.
\end{equation}

The CPL parameterization issue at high redshift $z$ has been discussed in \cite{ref71,ref73}. To point out this behavior, the authors \cite{ref71,ref73} 
proposed CPL parameterization where parameters of $\omega_{0}$ and $\omega_{a}$ have the same definitions as those defined for parameterization of 
the CPL. Another we consider that Jassal-Bagla-Padmanabhan (JBP) parameterization of DE as

\begin{equation}
\label{31}
\omega=\omega_{0}+\omega_{a}\frac{z}{(1+z)^2}
\end{equation}

proposed in \cite{ref43}, this model represents  a dark energy component in both low and high redshift regions with rapid variation at low 
$z$ where the parameterization of the CPL can not be generalized to the whole universe. Here, the parameters $ \omega_{0}$ and  $\omega_{a}$ 
have the same meanings as those defined for the above two models. Barboza-Alcaniz parameterization proposed in \cite{ref44}, that this model 
presents a step forward in redshift areas where the CPL parameterization can not be reached out to the whole history of the universe. It is useful 
the structure is given by:

\begin{equation}
\label{32}
\omega = \omega_{0 }+ \omega_{a} z\frac{1+z}{(1+z^2)}
\end{equation}
which is well-behaved at $z\to{-1}$.
The other parameterization of DE include PADE-I and PADE-II
The EoS parameter can be written in terms of $z$ as:

\begin{equation}
\label{33}
\omega = \frac{\omega_{0 }+ \omega_{a} z\frac{z}{(1+z)}}{1+\omega_{b}\frac{1+z}{(1+z)}} 
\end{equation}

here the EoS parameter with $\omega_{b}\neq {0}$ maintains a strategic distance from the dissimilarity at $a\to\infty$ 
(or proportionally at $z = - 1$) \cite{ref45}.

\[\omega_{z} = \begin{cases}
{\frac{\omega_{0 }+ \omega_{a}}{1+\omega_{b}}} ~~~~  for ~~~ a\to 0~~~~ (z\to \infty~~ early~ time)\\
{\omega_{0 }=0~~ for~~~ a=1 ~~ (z=0~~ present~~ time)}\\
{\frac{\omega_{0 }}{\omega_{b} }~~ for~~ a\to \infty ~~~(z\to -1~~ future~~ time)}\\
\end{cases}
\]

Here we have to set $\omega_{b}\neq {0}$ and ${-1}$. In this manner we conclude that 
the PADE (I) formula is a well-behaved function in the scope of 
$\leq{a}\leq\infty$ (or comparably at $-1 \leq{z}\leq\infty$). Clearly PADE (I) estimation has three free parameters: 
$\omega_{0},~ \omega_{a },$ and $\omega_{b}$, to evade singularities in the cosmic 
extension, $\omega_{b}$ necessities to lie in the range $-1 < \omega_{b} < 0$.\\
Here the present parameterization is composed as a component of $ ln a$. Right now, EoS 
parameter can be composed as
\begin{equation}
\label{34}
\omega = \frac{\omega_{0 }+ \omega_{a} ln\frac{1}{(1+z)}}{1+\omega_{b}ln\frac{1}{(1+z)}}, 
\end{equation}

where $\omega_{0}, \omega_{a },$ and $\omega_{b}$ are constants \cite{ref45} 
In PADE (II) parameterization, to keep away from singularities at these ages, we 
need to force $\omega_{b}\neq0$

\[\omega_{z} = \begin{cases}
{\frac{ \omega_{a}}{\omega_{b}}} ~~~~  for ~~ a\to 0 ~~~(z\to \infty~~ early~~ time)\\
{\omega_{0 }=0 ~~~ for~~ a=1 ~~ (z=0~~ present~~ time)}\\
{\frac{\omega_{a }}{\omega_{b} } ~~ for ~~ a\to \infty~~ (z\to -1~~ future ~~ time)}\\
\end{cases}
\]

	\begin{table}[H]
	\caption{Dark energy parameterization with best fit values of  $\omega_{0}$  and $\omega_{a}$.}
	\begin{center}
	\begin{tabular}{ |c|c|c|c |}
	\hline
	\tiny	Model  & \tiny  Parameterization & \tiny best fit parameter using SNe Ia LA &  \tiny Reference \\
	\hline
			
        \tiny  CPL & \tiny $\omega=\omega_{0}+ \omega_{a}\frac{z}{(1+z)}$ & \tiny $\omega_{0 }=-0.991\pm0.036~~\omega_{a}=0.297\pm 0.779$  
        &\tiny \cite{ref39},\cite{ref55}\\ 
	\hline	
			
	\tiny	$JBP$	& \tiny $\omega=\omega_{0}+\omega_{a}\frac{z}{(1+z)^2}$ &\tiny $\omega_{0 }=-1.013\pm0.070~~\omega_{a}=-0.297\pm 4.306$ 
	& \tiny \cite{ref43} \cite{ref55} \\ 
	\hline

	\tiny$BA$ & \tiny $\omega = \omega_{0 }+ \omega_{a} z\frac{1+z}{(1+z^2)}$  &\tiny $\omega_{0 }=-0.997\pm0.034~~\omega_{a}=-0.245\pm 0.545$ 
	&  \tiny \cite{ref44} \cite{ref55} \\
	\hline
			
	\tiny$PADE-I$	& \tiny $\omega = \frac{\omega_{0 }+ \omega_{a} z\frac{z}{(1+z)}}{1+\omega_{b}\frac{1+z}{(1+z)}} $ 
	&\tiny $\omega_{0 }=-0.825\pm0.091~~\omega_{a}=-0.683\pm 0.040 $ &\tiny \cite{ref45} \cite{ref55} \\
	\hline
	\tiny$PADE-II$	& \tiny $\omega = \frac{\omega_{0 }+ \omega_{a} ln\frac{1}{(1+z)}}{1+\omega_{b}ln\frac{1}{(1+z)}} $ 
	&\tiny $\omega_{0 }=-0.889\pm0.080~~\omega_{a}=0.297\pm 0.779$ &  \tiny \cite{ref45} \cite{ref55} \\
	\hline
	\end{tabular}
	\end{center}
        \end{table}

Figure 9 shows the behaviour of the EoS parameter verses redshift $z$ for five DE parameterization. All the parameterization, 
contain two parameters $\omega_{0 }$ and $\omega_{a}$ and the the parameterization (PADE-1,PADE-II) containts three parameters. 
Here $\omega_{0 }$  is related to the present value of the equation of state for the dark energy and $\omega_{a}$ $\&$  $\omega_{b}$ 
determines its evolution with time. In this framework, using the best-fit values, we found that only the PADE-II
 parameterization remains in the quintessence regime $(1 < \omega < 1/ 3)$ and at high redshifts, while they enter in the phantom regime. 
 The rest of the  parameterizations(PADE-1, CPL, BA, JBP) evolve in the phantom region $(\omega < -1)$ at high redshifts.

\begin{figure}[H]
	\centering
	\includegraphics[width=7cm,height=7cm,angle=0]{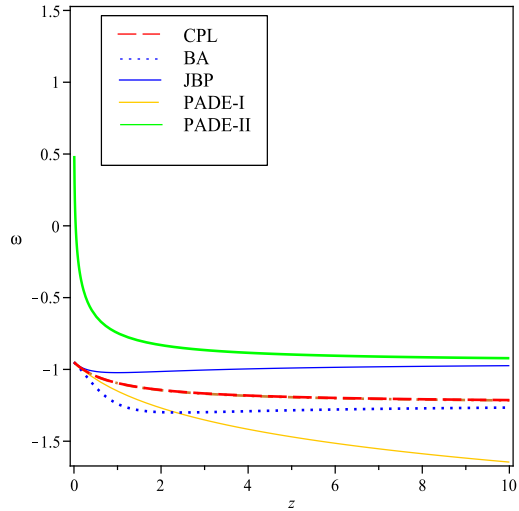}
	\caption{Plot of EoS parameter $\omega $ versus $z$ for five DE parameterization } 
\end{figure}
\section{Correspondence of Quintessence Dark Energy Model with Swampland Criteria }

Swampland criteria may also be examined for alternative Gravity theories \cite{ref83,ref84}. It is particularly
interesting because some modified theories are a useful paradigm to cure shortcomings of General Relativity at
ultraviolet and infrared scales, due to the lack of a full quantum gravity theory \cite{ref85}.
In this direction Bebetti and Capozzilla \cite{ref85} worked on the Noether Symmetry method with Swampland conjecture 
in f(R) gravity. The selected f(R) models, satisfying the swampland conjecture, are consistent, in principle, with both early 
and late-time cosmological behaviors. In this section we are discussing the consequences of swampland's DE principles in the 
framework of current and prospective cosmological observations. The swampland refers to a set of low-energy theories which look 
consistent from low- energy perspective but fail to be UV completed with quantum gravity. \\

 The swampland Conjecure concerning the effective potential $V (\phi)$ of the scalar field ( $\phi$)  are provided by subsequent inequalities:

\begin{equation}
\label{35}
Conjecure~~ {1}~~~~~:  ~~ M_{p1}\mid\triangledown V\mid \ge c_{1}V
\end{equation}
Here $M_{p1}$ is the reduced planck mass. The field should not, additionally, fluctuate more than around one planck unit over the entire 
universe's history.

\begin{equation}
\label{36}
Conjecure~~ {2}~~~~~:  ~~\triangle\phi \lesssim c_{2}M_{p1},
\end{equation}
where  $c_{1}$ and $c_{2}$ are the constants of first order has become significant,

or otherwise  light fields become important and the effective field theory becomes invalid. The swampland criteria have started 
expansions  proposed in \cite{ref65}. Recently lots of activity, due to their suggestions for inflation are discussed in 
\cite{ref86}-\cite{ref95}. \\

The gravitational field and the scalar field are depicted by the action	
		
\begin{equation}
\label{37}
S=\int d^4 x\sqrt{-g}\left(\frac{Mp^2}2{(R)} + L_{m}-\frac{1}2 g^{ij}\partial_{i}\phi \partial_{j}\phi-V(\phi)\right),
\end{equation}

where $R$ stands for Ricci scalar, and $V(\phi)$ is a potential and $\phi$ is the scalar field. \\

In this section we focus on the  exponential quintessence model with a potential of the form $ V(\phi)= V_{0} e^{\lambda_{1}\phi}$ \cite{ref93}
We compute the variations of dark energy equation of state 
for $\lambda_{1} = M_{p1}V^{'}/V$ and compare them with the dark energy parametrisation.  
The behavior for the the pressure and the energy density of the quintessence field \cite{ref96} is given as,
\begin{equation}
\label{38}
\rho_{\phi}=\frac{1}{2}\dot \phi^{2}+V(\phi) ,~~ p_{\phi}=\frac{1}{2}\dot \phi^{2}-V(\phi)
\end{equation}
Therefore the string theory criteria can be used to constrain the dark energy models that originate from a scalar field theory. 
Numerous  works in literature that address Swampland criteria and quintessence models \cite{ref87}-\cite{ref92}. The string Swampland 
criteria for an effective field theory to be consistent with the string theory. The quintessence models of dark energy were significantly 
compelled by the string theory of swampland criteria.\\

The behaviour of  $\phi$ and $V(\phi)$  are shown in Figs. $10$ and $11$. It is obvious that the scalar field $\phi$ is the 
increasing function verses $z$ and potential field $V(\phi)$  in the terms of the scalar field $\phi$ are also a positive decreasing 
function associated with swampland criteria 1 $\&$ 2.

The swampland criteria gives tight limitations to the dark energy models of the late time acceleration of the Universe just as on in 
stationary models of the early universe \cite{ref97}-\cite{ref99} right now, the  basis of present and future cosmological perceptions.

\begin{figure}[H]
	\centering
	\includegraphics[width=7cm,height=7cm,angle=0]{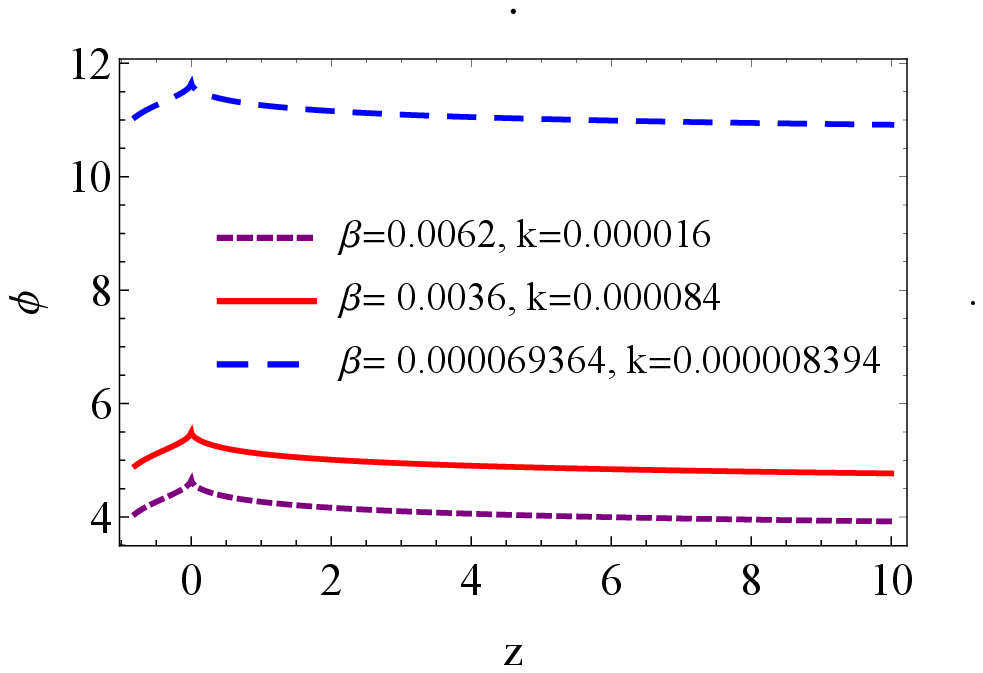}
	\caption{Plot of scalar field  $(\phi)$ versus $z$. Here $\lambda = 1$, $\alpha = 26.6$, $ m= 0.004$. } 
\end{figure}

\begin{figure}[H]
\centering
\includegraphics[width=7cm,height=7cm,angle=0]{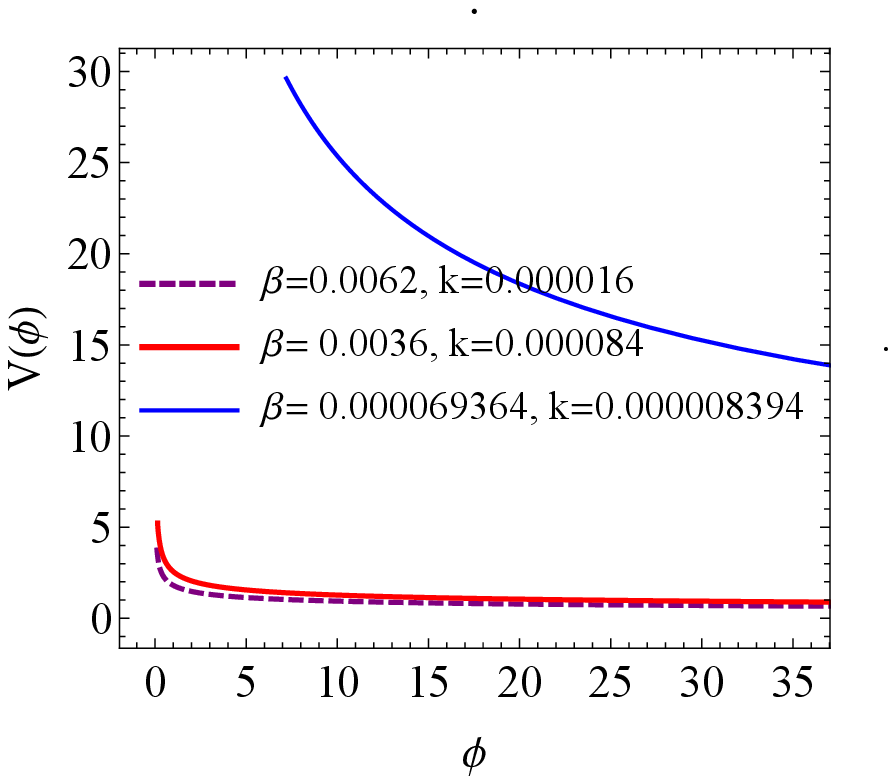}
\caption{Plot of $V(\phi)$ versus $\phi$. Here $\lambda = 1$, $\alpha = 26.6$, $ m= 0.004$.} 
\end{figure}

\begin{figure}[H]
	\centering
	\includegraphics[width=7cm,height=7cm,angle=0]{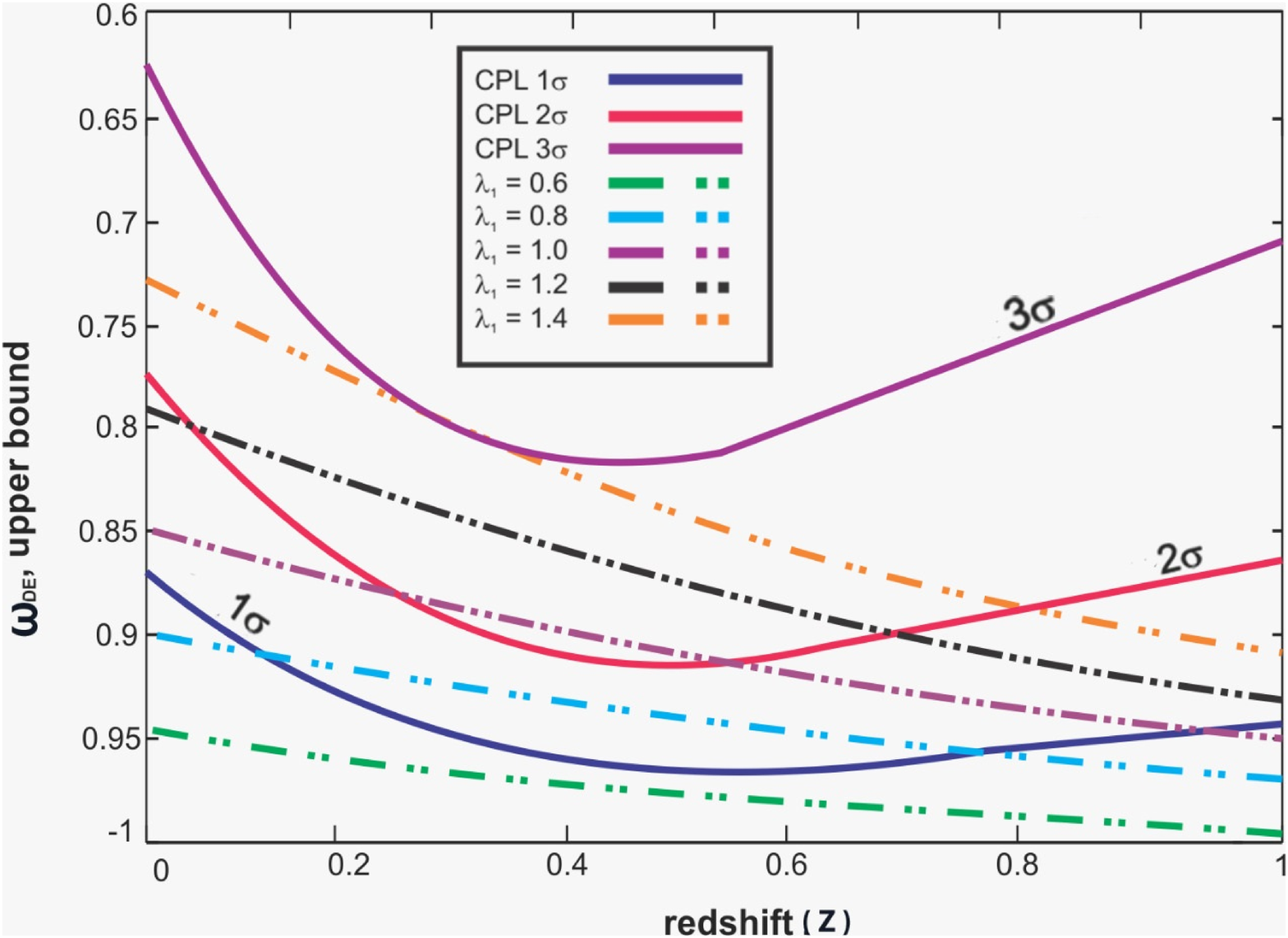}
	\caption{Plot of EoS $(\omega_{upper bound}) $ versus $z$ with CPL parameterization  . } 
\end{figure}

A discussion  on the significance of exponential type of potential to address the validity of swampland models is given in \cite{ref100}. 
Dark energy has likewise been considered in \cite{ref94} for dark energy equation of state investigation with CPL parametrization with 
regards to examining swampland measures. In  figure 12  we plot equation of state  as a function of red-shift. We further analyze our 
dark energy with the 1$\sigma$, 2$\sigma$, 3$\sigma$ upper limits of $\omega$, developed by CPL parameterization, with the recent 
cosmological results given in \cite{ref100}.
We are  calculating our results of $\omega$ vs z for standard quintessence dark energy model for values of $\lambda_{1}$ where 
$\lambda_{1} = 0.6,0.8,1.0,1.2,1.4$ . In this figure if $\lambda_{1}=0.8$ the quintessence DE model lies below in the 2$\sigma$ 
and 3$\sigma$ upper bounds and $\lambda_{1}=1, 1.6$ quintessence DE model lies below in the  3$\sigma$ upper bounds and and satisfied 
the Swampland criteria. Therefore the quintessence dark energy model is fulfills and satisfied  the  swampland criteria for recent data. 
Hence we have examined the results of swampland parameters on scalar DE field models, to be specific general core.

\section{Conclusions}
In this paper we have presented anisotropic Bianchi Type$-VI_{h}$ space-time filled with anisotropic fluid in the frame work of $f(R,T)$ 
gravity propose by \cite{ref35}. The motive is to obtain a set of field equations with a time-dependent $\Lambda$. \\

The main features of the models are as follows: \\

\begin{itemize}

\item We have discussed our theoretical models based on three data sets: (i) supernova type Ia union data \cite{ref49}, (ii) BAO and 
CMB data \cite{ref50} and (iii) current data in combination with HOD and LA observations \cite{ref51,ref52}.

\item The solution of the corresponding field equations is obtained by assuming a time-dependent DP $q(t)=-1+\frac{\beta}{\sqrt{2\beta t +k}}$, 
where $k$ and $\beta$ are arbitrary integrating constants. In plotting all figures, we have used three sets of ($\beta, k$): 
(i) $\beta = 0.0036$, $k = 0.000084$, (ii) $\beta = 0.0064$, $k = 0.000016$ and (iii) $\beta = 0.000008394$, $k = 0.0069364$ respectively (Table-1). 
These three sets of values of ($\beta, k$) are obtained by using three observational data sets \cite{ref49}-\cite{ref52}.

\item The transition redshift for our derived models for two cases (iii) $\beta = 0.000008394$, $k = 0.0069364$ and 
(ii) $\beta = 0.0062$, $k = 0.000016$ are found to be $z_{t} = 1.954$ and $z_{t} = 2.28$ respectively (Fig. $1(b)$) which are in good 
agreement with observational values \cite{ref101}-\cite{ref106}. The current $H(z)$ data redress a big redshift range, $0.7 \leq z \leq 2.36$ 
\cite{ref101}-\cite{ref106} larger than that covered by Type Ia supernovae ($0.01 \leq z \leq 1.30$). The $H(z)$ data can be used to trace the cosmological 
deceleration-acceleration transition \cite{ref60, ref107}.

\item For two cases (ii) $\beta = 0.0062$, $k = 0.000016$, and (iii) $\beta = 0.000008394$, $k = 0.0069364$, the DP varies from deceleration 
(early time) to acceleration (present time) whereas for another case (i) $\beta = 0.0036$, $k = 0.0000084$, we find only accelerating universe 
(Fig. $1(a)$).

\item For all the three cases, the energy density ($\rho$) is found to be a decreasing function of time and remains always positive in the 
whole evolution of the Universe (Fig. $2(a)$). The Fig. $2(b)$ shows the variation of $\rho$ with redshift ($z$). From this Fig. $2(b)$ 
we observe that $\rho$ is an increasing function of $z$ which is consistent with well established law.

\item The anisotropic fluid energy density ($\rho_{B}$) is declining time function and remains always positive (Figs. $3(a)$). We also observe that 
$\rho_{B}$ is an increasing function of redshift (Figs. $3(b)$) for all three cases.

\item The fluid pressure ($p$) is negative increasing function of cosmic time $t$ and remains always negative (Figs $4(a)$) which show the 
existence of dark energy (negative pressure). The $p$ is decreasing function of redshift (Figs $4(b)$).

\item We observe that for one case (ii) the universes are varying in quintessence $\omega\ge{-1}$ region \cite{ref108} through of evolution, while 
later on crosses PDL ($\omega = -1$) and finally approaches to phantom region $\omega\le{-1} $ \cite{ref109} (Figs. $5(a)$ \& $(b)$). In other 
two cases (i) \& (iii) the universe is varying only in phantom region. Therefore, we conclude that we are living in phantom scenario.

\item The latest cosmological observations in our derived model confirm the existence of a decaying vacuum energy density of $\Lambda(t)$.
These results on type Ia supernova's magnitude and redshift suggested that our universe could extend through the cosmological 
$\Lambda$-term with induced cosmological density. From Figs. $6(a)$ \& $(b)$, we observe that the $\Lambda(t)$ is a   
positive decreasing function of cosmic time ($t$) in all three cases models of the Universe which is 
consistent with observations.

\item We have found that all energy conditions i.e. WEC, SEC and DEC are not satisfied for all the universes for all three values of 
($k, \beta$) (see Figs (7a, b \& c)). This is consistent with dark energy scenario.

\item We have observed that speed of sound remains less than light velocity ($c= 1$) throughout the universe evolution $(Figs. 8)$ in the 
three cases (i), (ii) \& (iii). This prove the physical acceptability of our solutions.

\item The parameterization of DE EoS ($\omega$) is significant to evaluate the dynamical part of the models. In Figs. ($9$), some renowned 
EoS parameterization like CPL, JBP, BA, PADE-I and PADE-II have been graphed. The results of our analysis of DE parameterization with 
the best-fit values of $\omega_{0}$ and $\omega_{a}$ have been given in Table-2. Based on this analysis, we put constraints on the model
parameters and found that the expansion data. In the present framework, using best-fit values, we found that only PADE-II and JBP 
parameterization remains in the quintessence regime and the rest CPL, BA and PADE-I evolve in the phantom region.

\item We have discussed the outcomes of swampland's DE precepts in the reference of current cosmological observations. The behaviour of 
$V(\phi)$ and $\phi$ are shown in Figs. $10$ \& $11$ respectively. We observe that potential is positive declines from finite value and 
disappears at late time (present era). The scalar field $\phi$ increases with time and always positive. In  figure 12  we plot equation 
of state  as a function of red-shift, in this figure we have analyze the dark energy with the 1$\sigma$, 2$\sigma$, 3$\sigma$ upper limits 
of $\omega$, developed by CPL parameterization, with the recent cosmological results.

\end{itemize}

Thus, our newly constructed models and their solutions are physically acceptable. Therefore, for the better understanding of the characteristics 
of Bianchi type-$ VI_{h}$ cosmological models in our universe’s evolution within the framework of $f(R,T)$ gravity theory and confrontation 
observational data, may be helpful.
\section*{Acknowledgment}
A. Pradhan thanks to the Inter-University Centre for Astronomy \& Astrophysics (IUCAA), India for its supportive assistant under 
visting associateship program. The authors are grateful to Prof. Y. Akrami and Dr. Umesh K. Shrama for their fruitful comments on Section 6 
which improves the paper in present form.


\end{document}